\documentclass[sigplan,screen,10pt,dvipsnames,svgnames,x11names,table]{acmart}
\renewcommand\footnotetextcopyrightpermission[1]{}
\usepackage[T1]{fontenc}
\usepackage{fancyhdr}

\usepackage{booktabs} 
\usepackage{listings}
\usepackage{xspace}
\usepackage{url,color}
\usepackage{breakurl}

\usepackage{hyperref}

\usepackage{amsmath,amsfonts}
\usepackage{algorithmic}
\usepackage{graphicx}
\usepackage{textcomp}
\usepackage{xcolor}

\usepackage{booktabs} 
\usepackage{listings}
\usepackage{xspace}
\usepackage{flushend}
\usepackage{pifont}
\usepackage{tikz}

\newcommand{\projectname}{\textsc{secureTF}\xspace}

\usepackage{grumble}

\newcommand{\myparagraph}[1]{\smallskip \noindent{\bf {#1}.}}

\newcommand{\out}[1] {}

\newcounter{codeLineCntr}

\reversemarginpar
\newif\ifnotes
\notestrue

\newcommand{\punt}[1]{}

\renewcommand{\eqref}[1]{Equation~(\ref{eq:#1})}

\newcommand{\proc}[1]{\ifmmode\mbox{\textsc{#1}}\else\textsc{#1}\fi}

  \newcommand{\func}[1]{\ifmmode\mathrm{#1}\else\textrm{#1}fi} %

\newcounter{remark}[section]

\usepackage[inline]{enumitem}
\setlist{noitemsep,topsep=0pt,parsep=0pt,partopsep=0pt}

\usepackage{setspace}
\usepackage[margin=5pt,font={stretch=0.9}]{caption}

\pdfoutput=1

\makeatother
\usepackage{tcolorbox}
\title{\textsc{secureTF}: A Secure TensorFlow Framework}

\author{Do Le Quoc, Franz Gregor, Sergei Arnautov} 
\affiliation{\institution{TU Dresden, Scontain UG}}

\author{Roland Kunkel}
\affiliation{\institution{TU Dresden}}
\author{Pramod Bhatotia}
\affiliation{\institution{TU Munich}}
\author{Christof Fetzer}
\affiliation{\institution{TU Dresden, Scontain UG}}

\renewcommand{\shortauthors}{D. L. Quoc et al.}


\begin{abstract}

Data-driven intelligent applications in modern online services have become ubiquitous. These applications are usually hosted in the untrusted cloud computing infrastructure. This poses significant security risks since these applications rely on applying machine learning algorithms on  large datasets which may contain private and sensitive information. 

To tackle this challenge, we designed \projectname, a distributed secure machine learning framework based on Tensorflow for the untrusted cloud infrastructure. \projectname is a generic platform to support unmodified TensorFlow applications, while providing end-to-end security for the input data, ML model, and application code.  \projectname is built from ground-up based on the security properties provided by Trusted Execution Environments (TEEs). However, {\em it extends the trust of a volatile memory region (or {\em secure enclave}) provided by the single node TEE  to secure a distributed infrastructure required for supporting unmodified stateful machine learning applications running in the cloud}.  

The paper reports on our experiences about the system design choices and the system deployment in production use-cases. We conclude with the lessons learned based on the limitations of our commercially available platform, and discuss open research problems for the future work.  
\end{abstract}

\begin{document}
\thispagestyle{plain}
\thispagestyle{fancy}	
\maketitle
\errorstopmode

\section{Introduction} 
\label{sec:introduction}
Machine learning has become an increasingly popular approach for solving various practical problems in data-driven online services~\cite{taigman2014deepface, bennett2007netflix,foster2014machine, deepmind_health}. While these learning techniques based on private data {\em arguably} provide useful online services, they also pose serious security threats for the users.  Especially, when these modern online services use the third-party untrusted cloud infrastructure for deploying these computations.



In the untrusted computing infrastructure, an attacker can compromise the confidentiality and integrity of the computation.  Therefore,  the risk of security violations in untrusted infrastructure has increased significantly in the third-party cloud computing infrastructure~\cite{Santos2009}.  In fact, many studies show that software bugs, configuration errors, and security vulnerabilities pose a serious threat to computations in the cloud systems~\cite{Gunawi_bugs-in-the-cloud, Baumann2014, Santos2012}.  Furthermore, since the data is stored outside the control of the data owner, the third-party cloud platform provides an additional attack vector.  The clients currently have limited support to verify whether the third-party operator, even with good intentions, can handle the data with the stated security guarantees~\cite{pesos, Vahldiek-Oberwagner2015}.

To overcome the security risks in the cloud, our work focuses on securing machine learning computations in the untrusted computing infrastructure.  In this context, the existing techniques to secure machine learning applications are limiting in performance~\cite{graepel2012ml}, trade accuracy for security~\cite{du2003using} or support only data classification \cite{bost2015machine}. 
Therefore, {\em we want to build a secure machine learning framework that supports existing applications while retaining accuracy, supporting both training and classification, and without compromising the performance. Furthermore, our work strives to provide end-to-end security properties for the input data, ML models, and application code}.

%
%

To achieve our design goals,  trusted execution environments (TEEs), such as Intel SGX~\cite{intel-sgx} or ARM TrustZone~\cite{arm-trustzone}, provide an appealing way to build a secure machine learning system.  In fact, given the importance of security threats in the cloud, there is a recent surge in leveraging TEEs for shielded execution of applications in the untrusted infrastructure~\cite{Baumann2014,arnautov2016scone, tsai2017graphene, shinde2017panoply, Orenbach2017}. {\em Shielded execution} aims to provide strong confidentiality and integrity properties for applications using a hardware-protected secure memory region or {\em enclave}.


While these  shielded execution frameworks provide strong security guarantees against a powerful adversary, the TEEs have been designed to secure single-node in-memory (volatile) computations. Unfortunately, the trust of TEEs does not naturally extend to support distributed stateful applications running in the cloud. To build a secure machine learning framework that supports both training and classification phases, while providing all three important design properties: {\em transparency},  {\em accuracy}, and {\em performance}, we need to address several architectural challenges presented by TEEs, specifically Intel SGX, which acts as the root of trust. 

More specifically, in addition to  the conventional architectural challenges posed by the SGX architecture in the single node setting, such as limited enclave memory and I/O bottlenecks, we need to address the following three important challenges in the context of distributed cloud computing:

Firstly, we need to extend the trust of SGX to support the distributed TensorFlow framework, where the worker nodes are running in the remote distributed enclaves while ensuring that they execute correct code/computations and data. However, this is a challenging task since Intel SGX is designed for securing single machine computations. 

Secondly,  we need to support practical features offered by the virtualized platforms in the public cloud service to enable {\em elastic and fault-tolerant computing}, i.e., scaling-up/down based on the workloads, and dealing with failures/migrations. To support these important requirements, we need to ensure the new worker node running in a container preserves the integrity, confidentiality of the data, ML models, and application code. However, the traditional remote attestation  using the Intel Attestation Service (IAS)~\cite{costan2016intel} is impractical to support the elastic and fault-tolerant computing. Therefore, we need to redesign the mechanism to ensure an elastic trust establishment through a configuration and attestation service.

Lastly, we need to support  stateful machine learning applications that rely on  reading the input data or write computation results from/to a file system storage as well as to the network. Unfortunately, Intel SGX is designed to protect only the data and computation residing in the volatile enclave memory. It does not provide any security guarantees for stateful machine learning computations across multiple machines.

To overcome these design challenges, we present \projectname, a secure machine learning framework for the untrusted infrastructure. 
More specifically, we make the following contributions.
\begin{itemize}

\item We have designed and implemented \projectname as the end-to-end system based on TensorFlow  that allows secure execution of the existing unmodified TensorFlow applications without compromising the accuracy. 
 
\item We optimized the performance to overcome the architectural limitation of Intel SGX in the context of machine learning workloads for distributed untrusted cloud computing environments.

\item We report an extensive evaluation of \projectname based on microbenchmarks and production use-cases. Our evaluation shows that \projectname achieves reasonable performance overheads, while providing strong security with low TCB.

\end{itemize}

\noindent \projectname is a commercially available platform, and it is currently used in production by four major customers. In this paper, 
we report on our experiences on building  \projectname and deploying it in two production use-cases. We conclude the paper with the lessons learned based on the limitations of our system design, and a discussion on open research problems for the future work.

\section{Background and Threat Model}
\label{sec:background}

\subsection{Machine Learning using TensorFlow}

Machine learning aims to automatically extract useful patterns in large-scale data by building probabilistic models~\cite{simeone2017brief}. Machine learning approaches are often categorized into supervised, unsupervised and reinforcement learning.
All forms have in common that they require datasets, a defined objective, a model and a mechanism to update the model according to new inputs. 

To generalize the machine learning approach for masses, Google proposed TensorFlow~\cite{abadi2016tensorflow} as a machine learning framework designed for heterogeneous distributed systems. TensorFlow requires the user first to define a directed graph consisting of nodes representing operations on incoming data.
Nodes have zero or more inputs and outputs and perform operations on different levels of abstraction such as matrix multiplication, pooling or reading data from disk.
Nodes can also have an internal state, depending on their type.
Thus the whole graph can be stateful as well.

After defining the graph, the user can perform calculations by starting a session and running the previously defined operations.
TensorFlow uses a flow model for calculations.

Through the division of the calculation in the graph into nodes, TensorFlow makes it easy to distribute the execution across different devices. Therefore, TensorFlow can be deployed on mobile devices, single personal computers, as well as computer clusters, by mapping the computation graph on available hardware.

TensorFlow Lite~\cite{tensorflow-lite} is a feature-reduced version of TensorFlow, designed for mobile and embedded devices.
Optimization for mobile devices is achieved by running a mobile-optimized interpreter that keeps the load at a lower level and having the overall binary size smaller when compared to full TensorFlow.
The number of available operations for defining a graph is reduced to achieve a smaller memory footprint of the resulting binary.
This comes at the cost of trainability of the graph, because TensorFlow Lite can only perform forward passes in graphs. Instead, a model must first be training with the full version of TensorFlow and then exported and converted to a special TensorFlow Lite model format.
This format can then be used from the TensorFlow Lite API for inference.

\subsection{Intel SGX and Shielded Execution}
Intel Software Guard Extension (SGX) is a set of x86 ISA extensions for Trusted Execution Environment (TEE)~\cite{costan2016intel}. SGX provides an abstraction of a secure \emph{enclave}---a hardware-protected memory region for which the CPU guarantees the confidentiality and integrity of the data and code residing in the enclave memory. The enclave memory is located in the Enclave Page Cache (EPC)---a dedicated memory region protected by  an on-chip Memory Encryption Engine (MEE). The MEE encrypts and decrypts cache lines that are written and read to EPC, respectively. Intel SGX supports a call-gate mechanism to control entry and exit into the TEE. 

{\em Shielded execution} based on Intel SGX aims to provide strong confidentiality and integrity guarantees for applications deployed on an untrusted computing infrastructure~\cite{Baumann2014,arnautov2016scone, tsai2017graphene, shinde2017panoply, Orenbach2017}. Our work builds on the SCONE~\cite{arnautov2016scone} shielded execution framework.
In the SCONE framework, the applications are linked against a modified standard C library (SCONE libc). In this model, the application's address space is confined to the enclave memory, and interaction with the  untrusted memory is performed via the system call interface. In particular, SCONE runtime provides an {\em asynchronous system call} mechanism~\cite{flexsc}  in which threads outside the enclave asynchronously execute the system calls. 
Lastly, SCONE provides an integration to Docker for seamlessly deploying container images.

\subsection{Threat Model}

We  aim to protect against a very powerful adversary even in the presence of complex virtualization stacks in the cloud computing infrastructure~\cite{Baumann2014}. In this setting, the adversary can control the entire system software stack, including the OS or the hypervisor, and is able to launch physical attacks, such as performing memory probes.   In addition, we consider an untrusted network in the cloud environment, i.e., the adversary can drop, inject, replay, alter packages, or manipulate the routing of packages. This network model is consistent with the classic Dolev-Yao adversary model \cite{DolevYao}.
Even under this extreme threat model,  our goal is to guarantee the integrity, confidentiality, and freshness of data, code (e.g., Python code), and models of machine learning computation. 
We also provide bindings with Pesos~\cite{pesos}, a secure storage system to protect against rollback attacks~\cite{Parno2011} on the data stored beyond the secure enclave memory.
Our system is adaptable with SGXBounds~\cite{kuvaiskii2017sgxbounds}; therefore, \projectname is resilient to memory safety vulnerabilities~\cite{intel-mpx}.

However, we do not protect against side-channel attacks based on cache timing and speculative execution~\cite{foreshadow}, and memory access patterns~\cite{xu2015controlled, hahnel2017high}. Mitigating side-channel attacks is an active area of research~\cite{varys}. 
We do not consider denial of service attacks since these attacks are trivial for a third-party operator controlling the underlying infrastructure~\cite{Baumann2014}, e.g., operating system (OS), and hypervisor. 
Lastly, we assume that the  CPU hardware (including its implementation of SGX) are trusted and the adversary cannot physically open the processor packaging to extract secrets or corrupt the CPU system state.

\section{Design}
\label{subsec:overview}

In this section, we present the design of \projectname. 

\subsection{System Overview}
\begin{figure}
	\centering
	\includegraphics[width=0.4\textwidth]{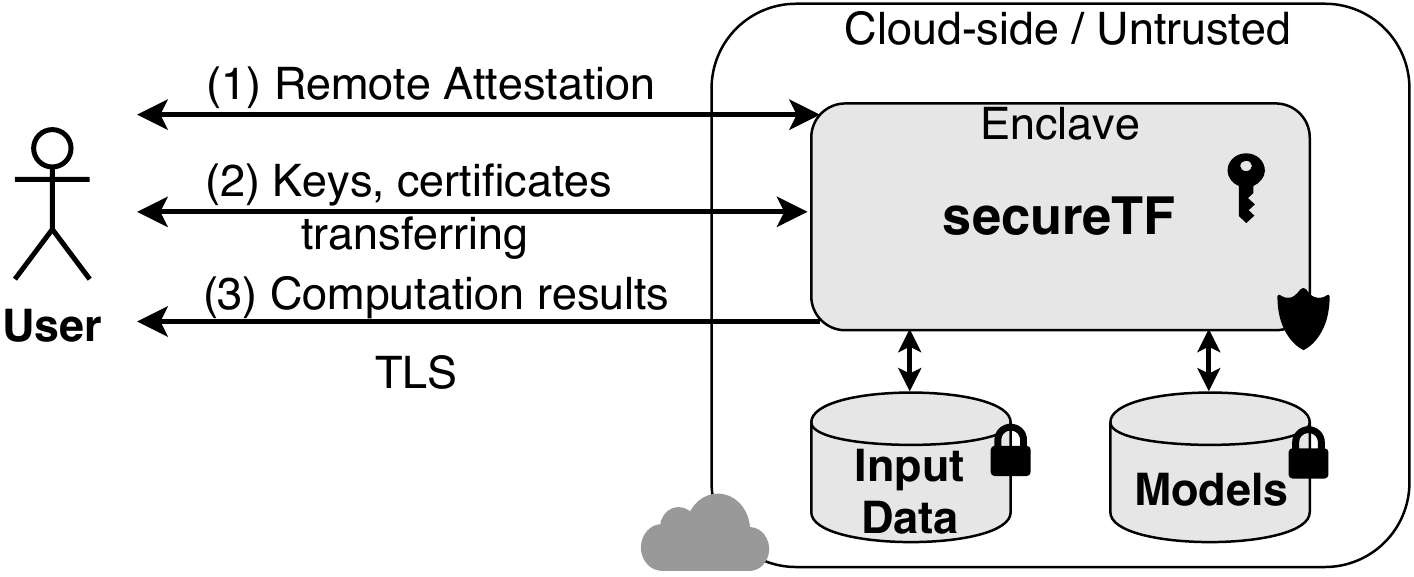}
	\caption{System overview.}
	\label{fig:overview}
\end{figure}

\projectname is designed for secure distributed machine learning computations using the hardware-assisted trusted execution environment (TEE) technologies such as Intel SGX. Figure~\ref{fig:overview}
depicts the high-level architecture of \projectname. Our system ensures not only the confidentiality, integrity and freshness of executions (e.g., training and classifying computations) but also the input data and machine learning models. At a high-level, the system works as follows: at the first step, when a user deploys a machine learning computation on a remote host (e.g., a public cloud), the user needs to establish trust into the \projectname instance running in the untrusted environment. To do so, the user performs the remote attestation mechanism provided by the TEE technology to ensure that the computation and the input data deployed in the remote environment are correct and not modified by anyone e.g., an attacker. After trusting the \projectname running in the remote environment, the user provides secrets including keys for encrypting/decrypting input and output data (e.g., input images and models, certificates for TLS connections), to the machine learning platform. After finishing the computation, \projectname returns the results back to the user via a TLS connection.

\myparagraph{Design goals}
Our primary design goal is to achieve strong confidentiality and integrity properties. By confidentiality, we mean that all data including models handled by the machine learning framework and the machine learning framework code itself may not be disclosed to or obtainable by an unauthorized party. By integrity, we mean that modifications of the data handled by \projectname that were done by an unauthorized party must be detectable and should not compromise the internal state and functioning. In addition, while designing a practical system, we aim to achieve the following goals.

%
%


\begin{itemize}
	\item{\em Transparency:}
	The secure  framework must offer the same interface as the unprotected framework, and should run unmodified existing applications based on TensorFlow. 
	
	\item{\em Performance:}
	We aim to impose as little overhead as possible when adding security to the machine learning framework.
	
	\item{\em Accuracy:} We do not aim to trade-off accuracy for security. Accuracy will be the same in the native TensorFlow framework as when using no security protection.
	
\end{itemize}

\subsection{Design Challenges}
\label{subsec:design-challenges}

Building a practical secure distributed machine learning system using TEEs such as Intel SGX is not straightforward, in fact, we need to handle several challenges.

\myparagraph{\raisebox{-1pt}{\ding{202}} Code modification} Intel SGX requires users to heavily modify the source code of their application to run inside enclaves. Thus, transparently supporting an unmodified machine learning framework to run inside enclaves is not a trivial task.

\myparagraph{\raisebox{-1pt}{\ding{203}} Limited EPC size}
Currently, Intel SGX supports only a limited memory space ($\sim94$MB) for applications running inside enclaves. However, most machine learning computations, especially training, are extremely memory-intensive.

\myparagraph{\raisebox{-1pt}{\ding{204}} Establishing the trust in a distributed system} 
Trust has to be established in the remote distributed enclaves to ensure that they execute correct code and data. However, this is a challenging task since Intel SGX is originally designed for a single machine.

\myparagraph{\raisebox{-1pt}{\ding{205}} Elastic and fault tolerant computing support}
Typically, public cloud services support {\em elastic computing}, i.e., when the input workload increases, the framework automatically spawns new service containers or instances to handle with the growth of requests. However, whenever spawning a new container, it requires to perform remote attestation to ensure the integrity, confidentiality of the machine learning application in that container before communicating with it. Unfortunately, the traditional attestation mechanism using the Intel Attestation Service (IAS)~\cite{costan2016intel} incurs significant overhead, thus it's impractical in this setting.

\myparagraph{\raisebox{-1pt}{\ding{206}} Stateful computing: security of network and file system}
Machine learning applications running inside SGX enclaves need to read input data or write results from/to a file system, storage systems, or network. Unfortunately, Intel SGX is designed to protect only the stateless in-memory data and computation residing inside enclaves. It does not provide any security guarantees for state stateful machine learning computations across multiple machines.

\subsection{System Design} 
\label{sec:design}

In this section, we present the detailed design of distributed \projectname that handles the aforementioned challenges in $\S$\ref{sec:introduction}.

\subsubsection{System Components}
\begin{figure}
	\centering
	\includegraphics[width=0.4\textwidth]{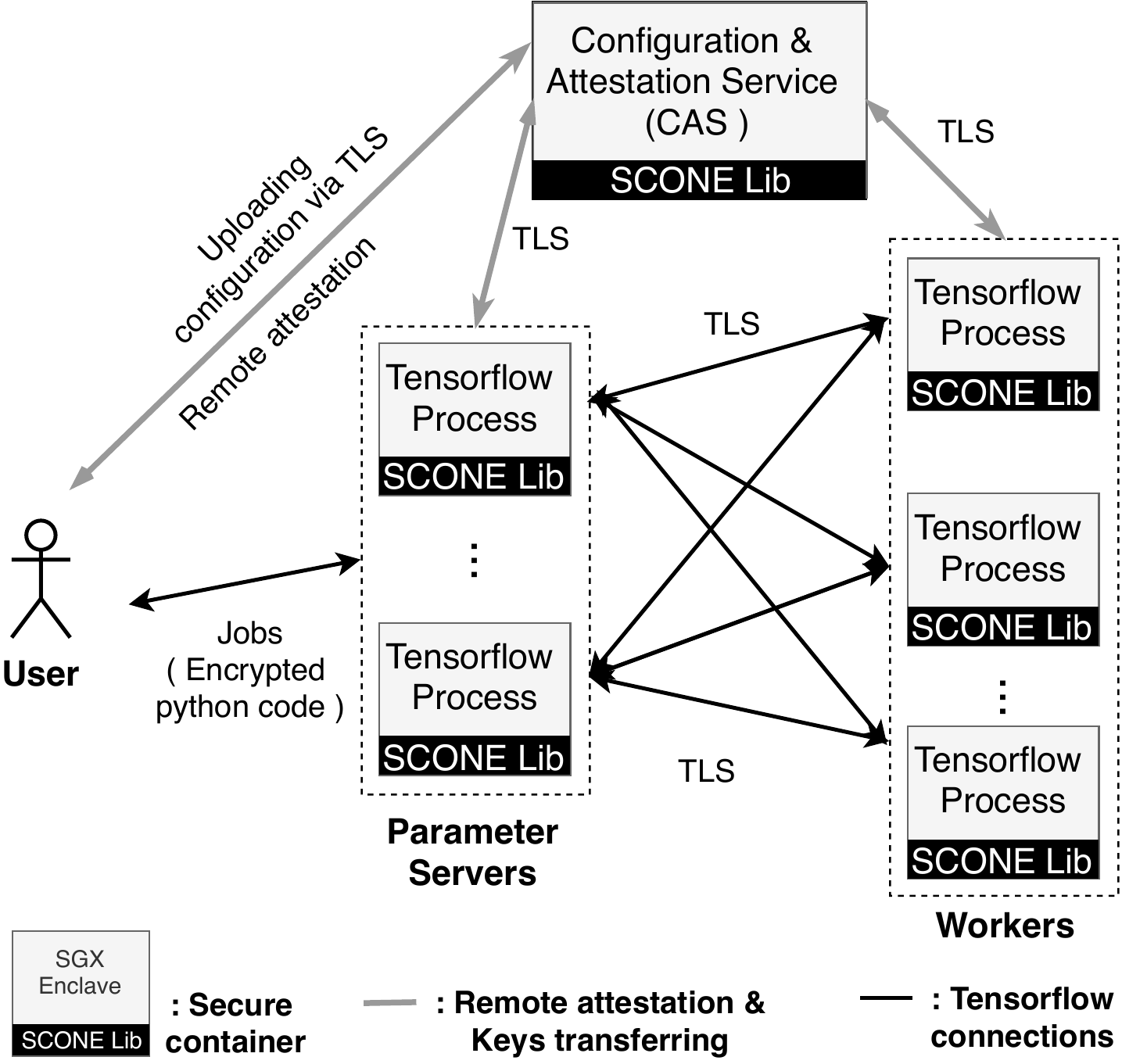}
	\caption{The distributed architecture of \projectname.}
	\label{fig:dist-tensorscone-arch}
\end{figure}

To overcome the challenge \raisebox{-1pt}{\ding{202}} (see $\S$\ref{sec:introduction}), we built \projectname based on the SCONE shielded execution framework~\cite{arnautov2016scone}. SCONE enables legacy applications to be executed in Intel SGX enclaves without source code changes. While there are other options available, we choose SCONE, because of the relatively small extra work required to run an application and comparatively small overhead compared to other available options. We leverage SCONE's Docker container support to design secure distributed \projectname which allows users to perform machine learning computations in a secure manner on an untrusted environment such as a public cloud. Figure~\ref{fig:dist-tensorscone-arch} shows the distributed architecture of \projectname. At the high-level, our systems consist of four core components: {\em Configuration and Remote Attestation Service (CAS)}, {\em secure machine learning containers} including Tensorflow parameter servers and Tensorflow workers, {\em network shield} and {\em file system shield}, and {\em adapted Tensorflow library}. 

We design the CAS component to handle the challenges \raisebox{-1pt}{\ding{204}} and \raisebox{-1pt}{\ding{205}}. This component takes an important role in the distributed architecture of \projectname which transparently and automatically performs the remote attestation for secure machine learning containers before transferring secrets and configuration needed to run them. The CAS component is deployed inside an SGX enclave of an untrusted server in the cloud or on a trusted server under the control of the user. When a secure machine learning container is launched, it receives the necessary secrets and configuration from CAS after the remote attestation process, to run machine learning computations using the adapted Tensorflow library running inside an enclave. 

We design the network shield and the file system shield components to address the challenge \raisebox{-1pt}{\ding{206}}.
All communications between secure machine learning containers and the CAS component are protected using the network shield component. 

Next, we provide the detailed design of each component. 

\subsubsection{Configuration and Remote Attestation Service}
CAS  enhances the Intel attestation service~\cite{costan2016intel} to bootstrap and establish trust across the \projectname containers and maintain a secure configuration of the distributed \projectname framework. 
CAS itself is deployed in an Intel SGX enclave. In the case that CAS is deployed in an enclave on an untrusted server, the user of \projectname needs to establish trust into the CAS instance, i.e., he/she needs to perform remote attestation of CAS before transferring encryption keys and certificates to process the encrypted input data and machine learning models. By using CAS, we can maintain the original distributed architecture of Tensorflow machine learning framework.

In addition, to guarantee the freshness of data during runtime, we design and implement an auditing service in CAS to keep track the data modification during machine learning computation. This mechanism allows \projectname to protect against rollback attacks.

\subsubsection{Secure Machine Learning Containers}
\begin{figure}
	\centering
	\includegraphics[width=0.32\textwidth]{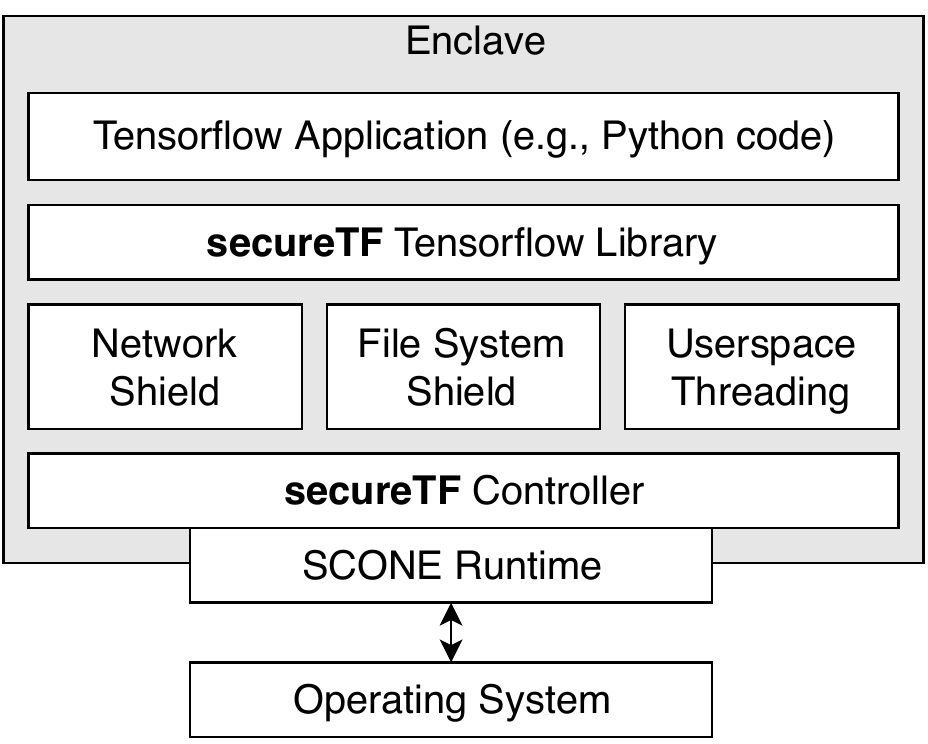}
	\caption[\projectname architecture]{
		The architecture of a secure machine learning container in \projectname.}
	\label{fig:tensorscone-arch}
\end{figure}

To build secure machine learning containers, we make use of TensorFlow and TensorFlow Lite. TensorFlow Lite has the additional advantage of having a smaller memory footprint which helps us to handle the design challenge \raisebox{-1pt}{\ding{203}}. We use SCONE~\cite{arnautov2016scone} as an additional layer that allows access to SGX features with fewer changes to application code. 
Figure ~\ref{fig:tensorscone-arch} presents the general architecture of a secure Tensorflow container using SCONE.

Since we build secure machine learning containers based on SCONE in \projectname, we use Docker~\cite{merkel2014docker} to conveniently deploy our system. No changes to the Docker engine is required. The design of a secure machine learning container in \projectname is composed of two components: (a) the \projectname controller that provides the necessary runtime environment for securing the TensorFlow library, and (b) \projectname TensorFlow library that allows deploying unmodified existing TensorFlow applications. We next describe these two components in detail.

\myparagraph{\projectname Controller} The \projectname controller is based on the SCONE runtime.
Inside the SGX enclave, the controller provides a runtime environment for TensorFlow, which includes the {\em network shield}, the {\em file system shield}, and the {\em user-level threading}. 
Data, that is handled through file descriptors, is transparently encrypted and authenticated through the shields.
The shields apply at each location where an application would usually trust the operating system, such as when using sockets or writing files to disk.
The shields perform sanity checks on data passed from operating system to enclave to prevent Iago attacks~\cite{Checkoway2013}.
More specifically, these checks include bound checks and checking for manipulated pointers.
This protection is required to fulfill the goal of not requiring the application to deal with untrusted systems (see challenge \raisebox{-1pt}{\ding{202}} in $\S$\ref{sec:introduction}).

\myparagraph{File system shield} The file system shield protects confidentiality and integrity of data files.
Whenever the application would write a file, the shield either encrypts and authenticates, simply authenticates or passes the file as is.
The choice depends on user-defined path prefixes, which are part of the configuration of an enclave.
The shield splits files into chunks that are then handled separately.
Metadata for these chunks is kept inside the enclave, meaning it is protected from manipulation.
The secrets used for these operations are different from the secrets used by the SGX implementation.
They are instead configuration parameters at the startup time of the enclave.

\myparagraph{Network shield}  TensorFlow applications do not inherently include end-to-end encryption for network traffic.
Users who want to add security must apply other means to secure the traffic, such as a proxy for the Transport Layer Security (TLS) protocol.
According to the threat model however, data may not leave the enclave unprotected, because the system software is not trusted.
Network communication must therefore always be end-to-end protected.
Our network shield wraps sockets, and all data passed to a socket will be processed by the network shield before given to system software.
The shield then transparently wraps the communication channel in a TLS connection on behalf of the user application.
The keys for TLS are saved in files and protected by the file system shield.

\myparagraph{User-level threading}
Enclave transitions are costly and should therefore be avoided when possible.
Many system calls require a thread to exit userspace and enter kernel space for processing.
To avoid thread transitions out of enclaves as much as possible, the controller implements user space threading.

When the OS assigns a thread to an enclave, it first executes an internal scheduler to decide, which application thread to execute. These application threads are then mapped to SGX thread control structures.
When an application thread blocks, the controller is run again to assign the OS thread to a different application thread instead of passing control back to the operating system.
In this way, the number of costly enclave transitions is reduced.
When no application thread is ready for execution, the OS either backs off and waits inside the enclave, or outside, depending on the time required for an enclave transition.
A side effect of this user-level threading scheme is that the controller does not require more OS threads than CPUs available to achieve full CPU utilization, which is usually the case for applications running under a conventional OS.

\subsubsection{\projectname TensorFlow Library}


Machine learning applications consist of two major steps.
In the first step, the model is trained, and thereafter, the model is employed for classification or inference tasks. Next, we explain the detailed design to run both training process and classification process with Intel SGX.

\myparagraph{Training process} For the training process, we use the full version of TensorFlow.
Training in TensorFlow is usually performed on acceleration hardware such as {GPU}s and distributed across multiple machines.
However, the \projectname controller requires SGX which is only available for CPUs. 
We therefore only support training on \textit{CPU}.
This limitation reduces the performance of the training process, but it is required to achieve the security goals.

The \projectname controller allows easy distribution of the application in the form of Docker images.
The training instances of \projectname can be distributed on multiple nodes, each running separate SGX hardware.
The network shield applies transparent protection of the communication channel between instances.
Scaling on the same instance, that is, on the same CPU is possible, but does decrease relative performance, because the limiting factor in our environment is EPC size, which is fixed for each CPU.
Only horizontal scaling with more nodes can substantially increase performance.


\myparagraph{Classification process} The main reason for dividing the classification and training process in our design is that we can use different TensorFlow variants for each step. Running with Intel SGX imposes less overhead, if applications have a smaller memory footprint, because the limited \textit{EPC} size is the major bottleneck (see challenge \raisebox{-1pt}{\ding{203}} in $\S$\ref{sec:introduction}).
TensorFlow Lite has a smaller memory footprint because it targets mobile devices.
The drawback is however that it cannot perform training by design.
Therefore, we can only use it for classification or inference. When protecting TensorFlow Lite with SCONE, the framework uses the SCONE C library instead of the common system library.
The internals of TensorFlow Lite do otherwise not require change, as the interface of the SCONE C library is compatible. The interface for using the classification method of \projectname is the same as for TensorFlow Lite.
Graph definitions created for TensorFlow Lite are compatible.


%

%

\section{Implementation} 
\label{sec:implementation}

We implement \projectname based on Tensorflow version $1.9.0$ and the SCONE framework~\cite{arnautov2016scone} to run machine learning computations within Intel SGX enclaves. We also consider other TEEs technologies such as ARM TrustZone~\cite{arm-trustzone} and AMD's TEE, SME/SEV~\cite{amd_secure_technology}. However, they have several limitations, e.g., ARM TrustZone supports only a single secure zone, and does not have any remote attestation mechanism, meanwhile, AMD's TEE does not support integrity protection~\cite{mofrad2018comparison}.

We rely on SCONE to implement some features such as file system shield and  network shield. However, it is not straightforward to use these features out-of-the-box to build \projectname. For example, SCONE does not support TLS connection via UDP which is required in Tensorflow. SCONE provides only confidentiality and integrity in network/storage shields, whereas, \projectname ensures also the freshness of data, code and models of machine learning computation.
In addition, the memory management and user-level multithreading need to adapt/extend it to fit the custom scheduler and memory management of TensorFlow framework. 
Thus, we need to develop these missing parts of these features to implement the design of \projectname.

In this section, we describe several challenges we faced during implementing \projectname and how we addressed them.
We first present how to enable the security features in \projectname during the training process ($\S$\ref{subsec:implement-training}) and classifying process ($\S$\ref{subsec:implement-classifying}). Thereafter, we describe the implementation of the CAS component in $\S$\ref{subsec:implement-cas}.


\subsection{Training Process}
\label{subsec:implement-training}
The typical user of TensorFlow uses the Python API for defining and training graphs, because it is the richest API.
Using Python with SCONE would impose additional complexity because it requires the \textbf{d}ynamic \textbf{l}ibrary \textbf{open} (\texttt{dlopen}) system call for imports.
As the name implies, \texttt{dlopen} dynamically loads libraries during runtime of a program.
However, \textit{SGX} does not allow an enclave to be entered by a thread, unless it has been finalized according to the procedures of enclave creation. 
A library that is dynamically loaded would therefore not be represented in the enclave's attestation hash. 
Consequently, \texttt{dlopen} is disabled by default for SCONE applications.
To allow {\em dlopen}, we need to change the SCONE environment accordingly (i.e., SCONE\_ALLOW\_DLOPEN=yes).
To ensure the security guarantee, we need to authenticate loaded libraries during runtime using the file system shield (see $\S$\ref{sec:design}). 

We support not only Python but also C++ API as native Tensorflow framework. In the previous version of \projectname, we did not support the Python API since, at that time, SCONE did not support {\em fork} system call which is required in the Python package~\cite{tensorscone-tech}. 
The C++ version covers the low-level API of TensorFlow, meaning many convenience features such as estimators or monitored training are not available. However, implementation using C++ API provides much better performance compared to using Python API.
There is one approach that let us use the convenience of the Python API for the definition of the computation graph.
TensorFlow allows exporting graphs and parameters, such as learned biases that were created in the current session.
Graph definitions and checkpoints containing the parameters can later be imported by another program.
Importing and exporting are available in both the C++ and the Python API, and they use interchangeable formats.
The user can therefore define a graph with the more high level Python API, including data inputs, and later import and run it with C++.
If the application does not by default already export its model with a named interface, changes are required to the original program, so that either the name of operations in the graph can be known, or an interface is defined.
%

%
For the training process, we used the full version of TensorFlow, not to be confused with TensorFlow Lite.
A graph definition must be provided by the user in form of a graph \textit{frozen} by a script packaged together with TensorFlow, when using either the Python or C++ API.
If the user has used the C++ API for the definition, the full source definition of the graph can also be used.

A frozen graph can be created from a graph definition exported from the Python script that defines the graph in the \textit{Protocol Buffers} (\cite{protobuf}) exchange format.
A checkpoint file containing all values of a graph that are not part of the graph definition, such as weights, biases and counters can be exported as well.

Alternatively, the graph can also be exported as a blank slate without any initialized internal values.
The initialization can then be done inside the \projectname environment, which is useful if a user wants to train the graph protected by SGX for the entire training process.
The initialization operations are required when using the Python API and are therefore usually part of the exported graph.

The user must also provide the inputs for training, such as a set of annotated images. \projectname protects the input data and code by activating the file system shield (see~$\S$\ref{sec:design}).

\subsection{Classification /Inference Process} 
\label{subsec:implement-classifying}
We implemented our design for inference/classifying computations in \projectname, by integrating the full Tensorflow with SCONE as we developed for the training computations.
In addition, we provide a light-weight version for inference by adapting Tensorflow Lite~\cite{tensorflow-lite} framework to run with SCONE.
We first ensured that Tensorflow and TensorFlow Lite compiles with the {\em musl} C library~\cite{alpine_faq} on Alpine Linux \cite{alpine_linux}, because SCONE enhanced the {\em musl} library to support legacy application running with Intel SGX.
The {\em musl} libc is designed to be compatible with The GNU C Library {\em (glibc)} but more secure with a smaller code base.
The issue we faced is that Tensorflow currently uses \textit{Identical code folding} (ICF)~\cite{ICF}, which is a compiler or linker feature, to eliminate identical function bodies at compile or link time in order to reduce the binary size.
However, it is currently supported by \textit{gcc} and the gold linker, but not by the {\em musl} linker or the compiler wrapper for {\em musl}.
We therefore removed the \textit{ICF} option for the binary targets in the TensorFlow source tree. Compiling the TensorFlow framework with and without ICF provides similar binary sizes.
Therefore, the performance cost when deactivating ICF will also be minimal.

The next issue is that TensorFlow also uses {\em backtrace} by default. This library is specific for{\em glibc}.
We therefore could not use it directly with musl. To solve this issue, we either recompiled dependencies against the musl libc, or disabled {\em backtrace} in the building configuration of Tensorflow. 



After adapting the Tensorflow source code, compiling it with SCONE is quite straightforward by merely setting the environment variables \texttt{CC} and \texttt{CXX} to the SCONE C and C++ compilers (i.e., \textbf{scone-gcc} and \textbf{scone-g++}).

Note that there is no standalone version of TensorFlow Lite available, meaning a user of TensorFlow Lite needs to build their application inside the TensorFlow source folder, with dependency targets set to TensorFlow Lite.
Tensorflow uses \textit{Bazel} as a build tool \cite{bazel}, however, Bazel also does not link library targets unless a binary target is created, which means TensorFlow Lite cannot be easily released from the source tree by compiling all libraries, and move them to the system's include directories.
Thus, we added compile targets that force linking as a workaround.
The libraries could then be moved to other projects along with the header files, and used as third party dependencies.
With this, we developed a classifier service  from scratch. The service takes classification requests via network, and uses TensorFlow Lite for inference/classifying.
For evaluation, we used an example available in the TensorFlow Lite source, which takes its inputs from the hard drive and prints the classification results to console.

\subsection{Configuration and Remote Attestation Service}
\label{subsec:implement-cas}
For large-scale deployment \projectname, we design the Configuration and Remote Attestation Service component (CAS) to transparently perform the remote attestation and transfer keys to distributed \projectname containers (see $\S$\ref{sec:design}).
We implement the CAS component using Rust~\cite{rust} programming language since it provides strong type safety. To run CAS with Intel SGX, we utilize the SCONE cross compiler to compile our implementation of CAS. 
We make use of an encrypted embedded SQLite~\cite{sqlite} to store encryption keys, certificates, and other secrets for Tensorflow computations (see $\S$\ref{sec:design}). This database itself also runs inside an enclave with the help of the SCONE framework.

To allow a user of CAS can ensure that the code of CAS was not modified and indeed runs inside a valid SGX enclave, besides running CAS with SCONE, we implement CAS in the way that it has zero configuration parameters that can control its behavior. Thus, an attacker with root/privileged accesses cannot break the trust given by the user in CAS. A detail description of CAS regarding protection against rollback attacks and key management is provided in~\cite{palaemon}.

\section{Evaluation}
\label{sec:evaluation}

In this section, we present the evaluation results of \projectname based on both microbenchmarks and macrobenchmarks with real world deployment.

\begin{figure}
	\centering
	\includegraphics[width=0.4\textwidth]{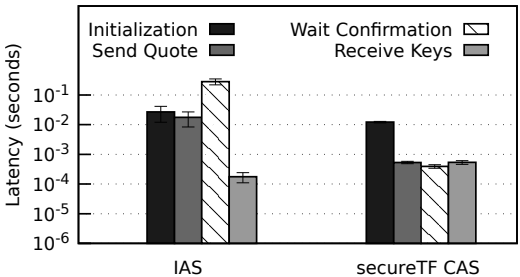}
	\caption{The attestation and keys transferring latency comparison between \projectname with the traditional way using IAS.}
	\label{fig:attestation}
	\vspace{-1mm}
\end{figure} 

\subsection{Experimental Setup} 
\label{subsec:eval-setup}
\myparagraph{Cluster setup} We used three servers with {em SGXv1} support running Ubuntu Linux with a 4.4.0 Linux kernel, equipped with an Intel\textcopyright{} Xeon\textcopyright{} \textit{CPU} E3-1280 v6 at 3.90\textit{GHz} with 32 KB L1, 256 KB L2, and 8 MB L3 caches, and 64 \textit{GB} main memory. These machines are connected with each other using a 1 Gb/s switched network. The CPUs update the latest microcode patch level.

In addition, we used a Fujitsu ESPRIMO P957/E90+ desktop machine with an Intel\textcopyright{} core
i7-6700 CPU with 4 cores at 3.40GHz and 8 hyper-threads (2 per core). Each core has a private 32KB L1 cache and a 256KB L2 cache while all cores share an 8MB L3 cache.


\myparagraph{Datasets}
We used two real world datasets: {\em (i)} Cifar-10 image dataset~\cite{krizhevsky2009learning} and {\em (ii)} MNIST handwritten digit dataset~\cite{mnist-dataset}. 

\begin{figure*}[t]
	\centering
	\includegraphics [width= 1\textwidth]{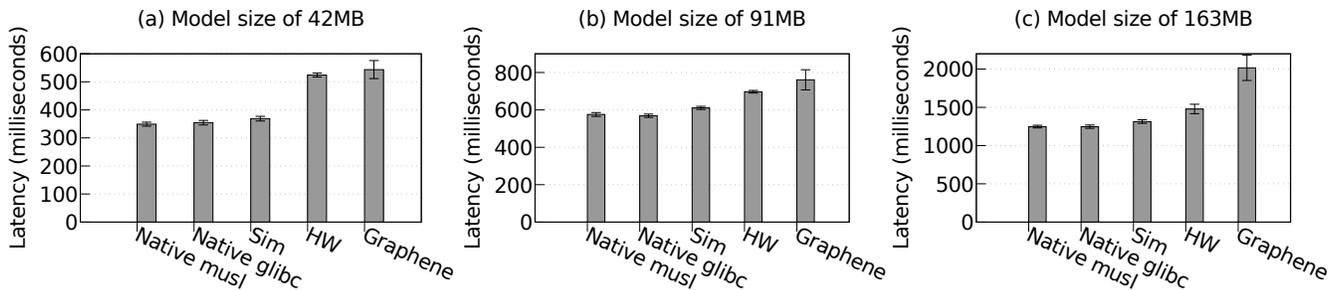}
	
	\caption{Comparison between \projectname, native versions and the state-of-the-art Graphene system in terms of latency with different model sizes, \textbf{(a)} Densenet (42MB), \textbf{(b)} Inception\_v3 (91MB), and \textbf{(c)} Inception\_v4 (163MB).} 
	\label{fig:classification-latency}
	\vspace{-1mm}
\end{figure*}

\myparagraph{\em \#1: Cifar-10} This dataset contains a labeled subset of a much larger set of small pictures of size 32x32 pixels collected from the Internet. It contains a total of 60,000 pictures.
Each picture belongs to one of ten classes, which are evenly distributed, making a total of 6,000 images per class.
All labels were manually set by human labelers. 
Cifar-10 has the distinct advantage that a reasonable good model can be trained in a relatively short time. The set is freely available for research purposes and has been extensively used for benchmarking machine learning techniques~\cite{xu2015empirical,he2016deep}.

\myparagraph{\em \#2: MNIST} The MNIST handwritten digit dataset\cite{mnist-dataset} consists of $60000$ $28\time28$ pixel images for training, and $10000$ examples for testing.

\myparagraph{Methodology}
Before the actual measurements, we warmed up the machine by running at full load with {\em IO} heavy operations that require swapping of {\em EPC} pages. We performed measurements for classification and training both with and without the file system shield. For full end-to-end protection, the file system shield was required.
We evaluate \projectname with the two modes: {\em (i)} hardware mode (HW) which runs with activated TEE hardware and {\em (ii)} simulation mode (SIM) which runs with simulation without Intel SGX hardware activated. We make use of this SIM mode during the evaluation to evaluate the performance overhead of the Intel SGX and to evaluate \projectname when the EPC size is getting large enough in the future CPU hardware devices. 

\subsection{Micro-benchmark: Remote Attestation and Keys Management}
In \projectname, we need to securely transfer certificates and keys to encrypt/decrypt the input data, models and the communication between worker nodes (in distributed training process). To achieve the security goal, we make use of the CAS component (see $\S$\ref{sec:design}) which attests Tensorflow processes running inside enclaves, before transparently provides the keys and certificates to encrypt/decrypt input data, models, and TLS communications. Note that the advantage of using CAS over the traditional way using IAS to perform attestation is that the CAS component is deployed on the local cluster where we deploy \projectname.

Figure~\ref{fig:attestation} shows the break-down latency in attestation and keys transferring of our component CAS and the method using IAS.
The quote verification process in our CAS takes less than $1$ms, whereas in the IAS based method is $\sim280$ms. In total, our attestation using CAS ($\sim17$ms) is roughly $19\times$ faster than the traditional attestation using IAS ($\sim325$ms). 
This is because the attestation using IAS requires providing and verifying the measured information contained in the quotes~\cite{costan2016intel} which needs several WAN communications to the IAS service. 

\begin{figure*}[t]
	\centering
	\includegraphics [width= 1\textwidth]{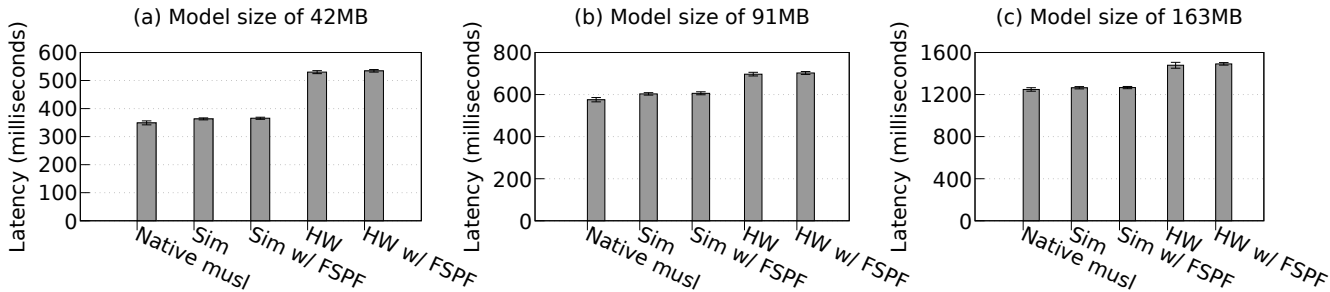}
	
	\caption{The effect of file system shield on the classification latency with different model sizes, \textbf{(a)} Densenet (42MB), \textbf{(b)} Inception\_v3 (91MB), and \textbf{(c)} Inception\_v4 (163MB).}
	\label{fig:classification-fspf}
	\vspace{-1mm}
\end{figure*}

\subsection{Macrobenchmark: Classifying Process} 
\label{subsec:eval-classifying}
We evaluate the performance of \projectname in real-world deployments. 
First, we present the evaluation results of \projectname in detecting objects in images and classifying images using pre-trained deep learning models. 
Thereafter, in the next section, we report the performance results of \projectname in training deep learning models (see $\S$\ref{subsec:eval-training}).

In the first experiment, we analyze the latency of \projectname in Sim mode and HW mode, and make a comparison with native versions using glibc and musl libc (i.e., running Tensorflow Lite with Ubuntu and Alpine linux) and a system~\cite{graphene-tesorflow-lite} provided by Intel using Graphene~\cite{tsai2017graphene}.  
Graphene is an open-source SGX implementation of the original Graphene library OS. It follows a similar principle to Haven~\cite{Baumann2014}, by running a complete library OS inside of SGX enclaves. Similar to SCONE~\cite{arnautov2016scone}, Graphene offers developers the option to run their applications with Intel SGX without requiring code modifications. All evaluated systems except the Graphene-based system run inside a Docker container.

To conduct this experiment, we use the Desktop machine (see $\S$~\ref{subsec:eval-setup}) to install Ubuntu 16.04 since the Graphene based system does not work with Ubuntu 18.04. 
The evaluated systems run with single thread because of the current version of the Graphene-based system does not support multiple threads, i.e., to run the classification process, we use the same input arguments for the classification command line: $\$\ label\_image -m\ model.tflite -i\ input.bmp -t\ 1$.
For the latency measurement, we calculate the average over $1,000$ runs. We use a single bitmap image from the Cifar-10 dataset as an input of evaluated systems. 

\myparagraph{Models}
For classifying images, we use several pre-trained deep learning models with different sizes including \textit{Inception-v3}~\cite{szegedy2016rethinking} with the size of $91$MB, \textit{Inception-v4}~\cite{szegedy2017inception} with the size of $163$MB and Densenet~\cite{huang2017densely} with the size of $42$MB.
We manually checked the correctness of a single classification by classifying the image with the TensorFlow {\em label\_image} application involving no self-written code and running directly on the host without containerization.
We later compared the results to the ones provided by \projectname and other evaluated systems, we could confirm that indeed the same classifying result was produced by the evaluated systems.

\myparagraph{\#1: Effect of input model sizes}
Figure~\ref{fig:classification-latency} shows the latency comparison between \projectname with Sim and HW mode, native Tensorflow Lite with glibc, native Tensorflow Lite with musl libc, and Graphene-based system. 
\projectname with Sim mode incurs only $\sim5\%$ overhead compared to the native versions with different model sizes. In addition, \projectname with Sim mode achieves a latency $1.39\times$, $1.14\times$, and $1.12\times$ lower than \projectname with HW mode with the model size of $42$MB, $91$MB, and $162$MB, respectively.
This means that operations in the libc of \projectname introduce a lightweight overhead. 
This is because \projectname handles certain system calls inside the enclave and does not need to exit to the kernel. In the Sim mode, the execution is not performed inside hardware SGX enclaves, but \projectname still handles some system calls in userspace, which can positively affect performance. We perform an analysis using {\em strace} tool to confirm that some of the most costly system calls of \projectname are indeed system calls that are handled internally by the SCONE runtime. 

Interestingly, the native Tensorflow Lite running with glibc is the same or slightly faster compared to the version with musl libc. The reason for this is that both C libraries excel in different areas, but glibc has the edge over musl in most areas, according to microbenchmarks \cite{clib_compare}, because glibc is tailored for performance, whereas musl is geared towards small size. Because of this difference in goals, an application may be faster with musl or glibc, depending on the performance bottlenecks that limit the application. Differences in performance of both C libraries must therefore be expected.

In comparison to Graphene-based system, \projectname with HW mode is faster and faster than Graphene-based system when we increase the size of input models, specially when it exceeds the limit of the Intel SGX EPC size ($\sim94$MB). In particular, with the model size of $42$MB, \projectname with HW mode is only $1.03\times$ faster compared to Graphene-based system, however, with the model size of $163$MB, \projectname with HW mode is $\sim1.4\times$ compared to Graphene-based system. The reason for this is that when the application allocates memory size larger than the EPC size limit, the performance of reads and writes severely degraded because it performs encrypting data and paging operations which are very costly. To reduce this overhead, we reduce the size of our libraries loaded into SGX enclaves. Instead of adding the whole OS libc into SGX enclaves as Graphene did, we make use of SCONE libc~\cite{arnautov2016scone} which is a modification of musl libc having much smaller size. In this library, system calls are not executed directly but instead are forwarded to the outside of an enclave via the asynchronous system call interface (see $\S$\ref{sec:design}). This interface together with the user level scheduling allows \projectname to mask system call latency by switching to other application threads. Thus,
we expect this speedup factor of \projectname compared to Graphene-based system  will increase more when the size of the input model size is increased and when the application runs with multiple threads.

\myparagraph{\#2: Effect of file system shield}
One of real world usecases of \projectname is that a user not only wants to acquire classifying results but also wants to ensure the confidentiality of the input images since they may contain sensitive information, e.g., handwritten document images. At the same time, the user wants to protect her/his machine learning models since he/she had to spend a lot of time and cost to train the models. To achieve this level of security, the user activates the file system shield of \projectname which allows he/she to encrypt the input including images and models and decrypt and process them within an SGX enclave (see $\S$\ref{sec:design}).

\begin{figure}[t]
	\centering
	\includegraphics [width=0.43\textwidth]{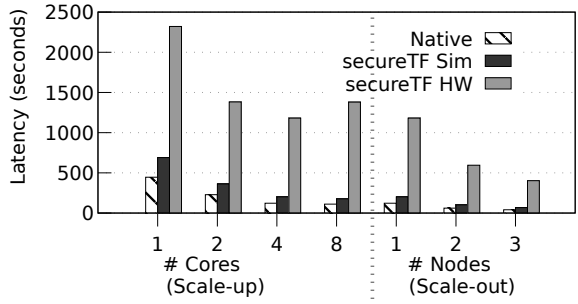}
	\caption{The latency comparison in classifying cifar-10 images with different numbers of CPU cores and nodes.}
	\vspace{-1mm}
	\label{fig:classification-scalability}
\end{figure}

In this experiment, we evaluate the effect of this file system shield on the overall performance of \projectname.
As previous experiments, we use the same input Cifar-10 images.
Figure~\ref{fig:classification-fspf} shows the latency of \projectname when running with/without activating the file system shield with different models. The file system shield incurs significantly small overhead on the performance of the classification process. \projectname with Sim mode running with the file system shield is $0.12\%$ slower than \projectname with Sim mode running without the file system shield. Whereas in the \projectname with HW mode, the overhead is $0.9\%$.
The lightweight overhead comes from the fact that our file system shield uses Intel-CPU-specific hardware instructions to perform cryptographic operations and these instructions can reach a throughput of up to 4 GB/s, while the model is about 163 MB in size. This leads to a negligible overhead on the startup of the application only. 

\begin{figure}
	\centering
	\includegraphics[width=0.42\textwidth]{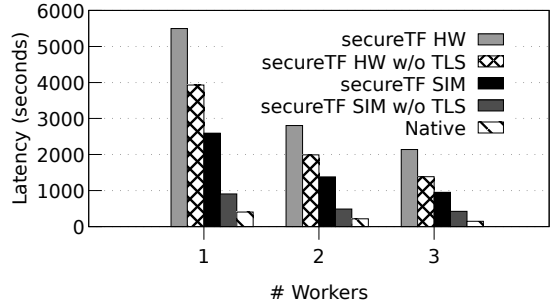}
	\caption{The training latency comparison between \projectname with different modes and native Tensorflow.}
	\label{fig:dist-traning-latency}
	\vspace{-1mm}
\end{figure} 

\myparagraph{\#3: Scalability}
To evaluate the scalability of \projectname, we measure the latency of \projectname in classifying $800$ cifar-10 images,  with different number of CPU cores (scale-up), and different number of physical nodes (scale-out). Figure \ref{fig:classification-scalability} shows that \projectname both in Sim and HW mode scale quite well from $1$ CPU core to $4$ CPU cores. 
However, \projectname in HW mode does not scale from $4$ CPU cores to $8$ CPU cores. The reason for this is that the EPC size is limited to  \textasciitilde$94MB$. When \projectname runs with $8$ cores it requires more than the capacity of the current version of Intel SGX. Thus, it requires to perform the paging mechanism which is very expensive. 
For scale-out evaluating, we keep each node to run with $4$ CPU cores. As we expected, \projectname in both Sim and HW mode scale well with different numbers of physical nodes. The latency of \projectname in HW mode with $1$ node is $1180$s whereas with $3$ nodes the latency is $403$s. 

\myparagraph{\#4: TensorFlow and TensorFlow Lite comparison}
To show the advantage of using TensorFlow Lite in \projectname instead of TensorFlow for inference or classification, we make a comparison between them. In this experiment, we use the same input model (i.e., Inception\_v3  model) and input image to evaluate the performance of \projectname using TensorFlow and TensorFlow Lite in HW mode. \projectname with TensorFlow Lite  achieves a $\sim71\times$ lower latency  ($0.697$s) compared to \projectname with TensorFlow ($49.782$s). The reason for this is that, the binary size of \projectname with TensorFlow Lite is only $1.9$MB, meanwhile the binary size of \projectname with TensorFlow is $87.4$MB; and note that the Intel SGX enclave EPC size is limited to $\sim94$MB.

\subsection{Macrobenchmark: Distributed Training}
\label{subsec:eval-training}

In this section, we evaluate the performance of \projectname in training distributed deep learning models at scale.
In these experiments, we use MNIST handwritten digit dataset (see $\S$\ref{subsec:eval-setup}) and three physical servers having the same configuration described in $\S$\ref{subsec:eval-setup}.
We keep the same batch size of $100$ and learning rate as $0.0005$, then measure the end-to-end latency of \projectname with different modes including HW mode, Sim mode, with and without activating the network shield, and a native version of Tensorflow. 

Figure~\ref{fig:dist-traning-latency} shows that \projectname with different modes scales almost linearly with the number of workers. \projectname, with full features running in HW mode, achieves a speedup of $1.96\times$ and $2.57\times$ when it runs with 2 and 3 workers, respectively. Unsurprisingly, this mode of \projectname is roughly $14\times$ slower compared to the native version due to the fact that the training process requires memory-intensive computations and the enclave EPC size is limited to $\sim94$MB. However, we believe that Intel will release new generation of its hardware which supports much large EPC sizes, thus we performed the experiments to evaluate \projectname in the SIM mode, to see the overhead of \projectname in the case the EPC size is enough for the training process. The slowdown factor in comparison to the native version, is reduced to $6\times$ and $2.3\times$ with \projectname in the SIM mode with and without activating the network shield, respectively. This indicates that the main overhead of the current implementation is the network shield. In addition, note that the slowdown in SIM mode is because of a scheduling issue in SCONE. We have reported this issue, it's now fixed in the current version of SCONE.

From the results of experiments, we can learn that with the current Intel SGX hardware capacity, performing securely inference/classification inside Intel SGX is practical, but it is not feasible for securely training deep learning (see $\S$\ref{subsec:lessons-learned}). 

\section{Real-World Deployments}
\label{sec:production}

\projectname is a commercial platform, and it is actively used by four customers (names omitted) in production.
We next present the  \projectname deployment for two use cases.

\begin{figure}
	\centering
	\includegraphics[width=0.36\textwidth]{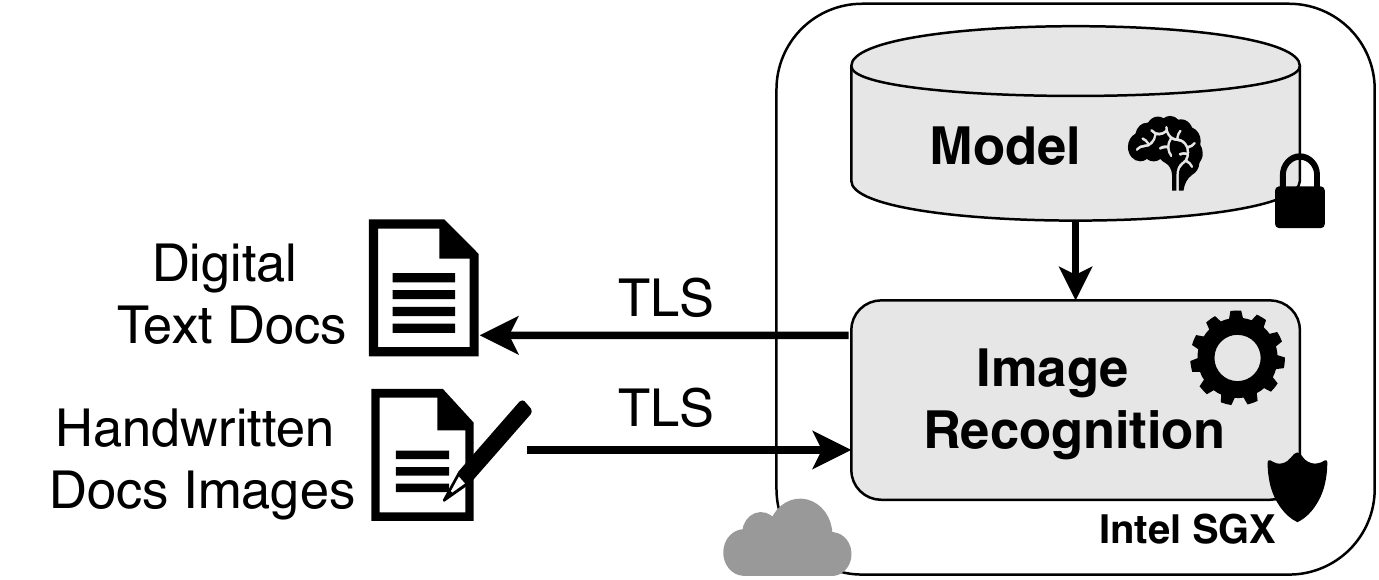}
	\caption{Deployment \#1: secure document digitization.}
	\vspace{-1mm}
	\label{fig:handwritting-usecase}
\end{figure} 

\subsection{Secure  Handwritten Documents Analysis}
\label{subsec:usecase1}

The first use case of \projectname is to perform secure handwritten documents analytics  (see Figure~\ref{fig:handwritting-usecase}). 
A company (name omitted) is using a public cloud to automatically translate handwritten documents into digital format using machine learning. Customers of this company not only want to acquire the inference results, but also want to ensure the confidentiality of the input since the handwritten document images contain sensitive information. At the same time, the company wants to protect its Python code for the inference engine as well as its machine learning models. 
To achieve this level of security, the company has deployed our framework---\projectname. The company uses the file system shield to encrypt Python code and models used for the inference. Meanwhile, the customers make use of the attestation mechanism of \projectname to attest the enclave running the service, and then send their handwritten document images via the TLS connections to this service to convert them into digital text documents.

\begin{figure}
	\centering
	\includegraphics[width=0.45\textwidth]{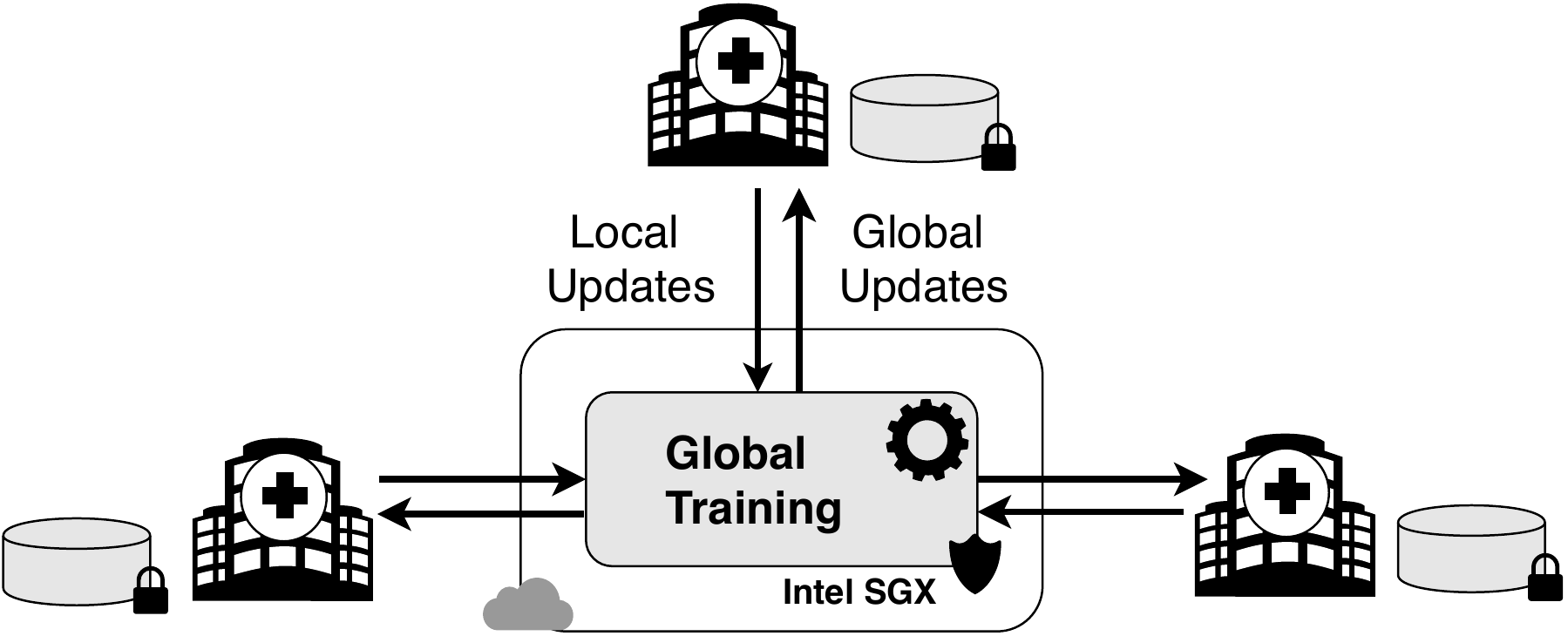}
	\caption{Deployment \#2: secure federated learning.}
	\label{fig:federatedlearning-usecase}
	\vspace{-1mm}
\end{figure} 

\subsection{Secure  Federated Learning: Medical Use-case}
\label{subsec:usecase2}

The second use case of \projectname is secure federated learning (FL)~\cite{federated-learning} (see Figure~\ref{fig:federatedlearning-usecase}). FL is proposed to allow multiple parties to jointly train a model that takes benefits from diverse datasets from the parties. In our second use-case, several hospitals are actively collaborating to train a model for  diagnosing brain tumors. However, at the same time, they want to protect patients' data regarding their privacy.  Thus, each hospital performs the training locally using its local data and thereafter shares the model parameters with the global training computation without sharing its actual data. Unfortunately, the local model may reveal private information~\cite{deeplearning-DP}.  These local models have been demonstrated to be vulnerable to several privacy attacks~\cite{federated-learning-privacy}. In addition, there is empirical evidence to the risks presented by machine learning models, e.g., the work by Matt \textit{et al} ~\cite{model-inversion-attacks} demonstrates that extracted images from a face recognition system look similar to images from the underlying training dataset.  To handle this issue, these hospitals make use of \projectname to run the global training inside Intel SGX enclaves. They only share their local model after performing the attestation over the enclaves. The communication with the global training enclaves are performed via TLS connections. 

\section{Discussion and Lessons Learned} 
\label{sec:discussion}

In this section, we discuss the lessons learned based on the limitations of our commercially available platform, and also, present open research problems for the future work.

\subsection{Training Vs Classification}
\label{subsec:lessons-learned}

The limited EPC size has different implications for training vs classifications. As shown in $\S$\ref{sec:evaluation},  training deep learning with the larger datasets inside the enclave is performance-wise limiting due to EPC paging. However, the EPC size is quite practical for classifying/inference processes since the size of the deployed ML model is usually much smaller than  the original training  data. As discussed in $\S$\ref{sec:production}, we are effectively using \projectname for image classification ($\S$\ref{subsec:usecase1}), and  federated machine learning use case (see $\S$\ref{subsec:usecase2}). 

To improve the performance of the training phase in the limited enclave memory regions, we are exploring two avenues: (1) {\em data normalization:} we can further improve the training performance, by normalizing input data, e.g., in image recognition services, all input images can be normalized to the size of $32\times32$; and  (2) {\em Ice lake CPUs} Intel announced the next generation processors called {\em ice lake} which supports larger EPC size~\cite{icelake}. 

\subsection{ML Model Optimizations}
To further improve the performance, we are exploring perform optimizations for the ML models leveraging  pruning and quantization tools, such as Intel OpenVINO Toolkit~\cite{openvino}. Since TensorFlow models are typically abstracted as directed computation graphs (see $\S$\ref{sec:background}), where nodes are operations and edges present the communication between operations. By performing optimization on the model graphs such as pruning unnecessary edges and nodes, we can significantly improve the performance of classification/inference computations.  The optimization also provides an opportunity to deploy ML inference service at edge devices supporting SGX~\cite{nuc} in edge computing. In fact, we have been working with an IoT-based company to use \projectname for securely deploying the latest trained models at the edge, while achieving high-performance.

\subsection{Security Analysis and Properties}
\label{subsec:secure analysis}
\projectname protects machine learning computations against attackers with privileged access by executing securely these computations inside Intel SGX enclaves. All data (input training/inference data, model, and Python code) and communications outside enclaves are  always encrypted. The encrypted data is only decrypted inside enclaves. The keys or secrets to decrypt the data are protected inside the CAS component which is also running inside an enclave. The CAS component only provides these secrets via TLS connections to the machine learning enclaves after attesting these enclaves.  A detailed security analysis of CAS is provided in~\cite{palaemon}. 
 
Intel SGX is typically vulnerable to side-channel attacks~\cite{sidechannel-attack1, sidechannel-attack2, gotzfried2017cache, vanbulck2018foreshadow, weisse2018foreshadowNG, spectre1, spectre2}. Although this type of attacks are out-of-scope of our work, it is worth to mention that the version of SCONE, which was integrated in \projectname, can not only protect against L1-based side channels attacks~\cite{varys} but also Iago attacks~\cite{checkoway2013iago}. We can also make use of LLVM-extensions, e.g., speculative load hardening~\cite{SpecLH2019} to prevent exploitable speculation which helps us to present the variants of Spectre attacks~\cite{spectre1, spectre2}.  In addition, the next generation of Intel CPUs~\cite{icelake} seems to provide hardware-based solutions to handle several types of side-channel attacks. 

\projectname supports only TLS-based communications to protect against eavesdropping on any communication between the CAS and  computation nodes in a distributed setting.
In \projectname, the TLS certificates are generated inside the SGX enclave running CAS, and thus they cannot be seen by any human. This mechanism allows \projectname to handle man-in-the-middle attacks. However, TLS and its predecessor are also vulnerable to side-channel attacks, e.g., attacks on RSA~\cite{drown-attack, robot-attack}. Thus, in \projectname, we recommend to completely disable RSA encryption and replace it by forward-secret key exchanges e.g., Elliptic-curve Diffie–Hellman (ECDHE) encryption~\cite{tls-book}.

\subsection{GPUs Support}
\label{subsec:gpu-support}
Graphics Processing Units (GPUs) have become popular and essential accelerators for machine learning~\cite{bekkerman2011scaling}. Unfortunately, trusted computing in GPUs is not commercially available, except research prototypes, such as Graviton~\cite{graviton}. 
Therefore, \projectname  provides security properties by relying on Intel SGX which is supported only for CPUs.

Technically, \projectname can also offer the GPU support, however, it requires weakening the threat model, i.e., we need to assume that the GPU computations and the communication between GPU and CPU are secure. The relaxation of the threat model may be acceptable in practice for several use cases, e.g., when users just want to protect their Python code and models for machine learning computations. \projectname can ensure the code and models are encrypted. However, this extension may not practical for many other use cases~\cite{graviton}. Therefore, we are currently  investigating GPU enclave research proposals, e.g., Graviton~\cite{graviton} and HIX~\cite{hix} which proposed hardware extensions to provide a secure  environment on GPUs. 

\section{Related Work}
\label{sec:related}

In this section, we summarize the related work about secure machine learning and shielded execution using Intel SGX. 

Early works on preserving-privacy data mining techniques have relied on randomizing user data \cite{du2003using, bost2015machine, PrivApprox2017}.
These approaches trade accuracy for privacy. They include a parameter that allows making a trade-off between privacy and accuracy.
The proposed algorithms aim to provide privacy of computation, but they do not protect the results themselves in the cloud, nor do they secure the classification phase. 
While this can protect the users privacy, it does not cover training as in \projectname. Further, we target to provide the same accuracy level as the native execution.

Graepel et al. \cite{graepel2012ml} developed machine learning algorithms to perform both training and classification on encrypted data.
The solution is based on the properties of homomorphic encryption.
However, homomorphic encryption schemes provide restrictive computing operations, and incur high performance overheads. 
There have been a series of recent works~\cite{secureml, gazelle, cryptflow, delphi} aimed to provide secure machine learning platforms with Secure multiparty computation (MPC). Especially, Delphi~\cite{delphi} and CrypTFlow~\cite{cryptflow} demonstrated that they outperform previous works. However, these systems also were designed only for securing inferences.  \projectname is instead based on a hardware-based encryption approach (i.e., Intel SGX) and it supports both training and inference computations.

Shielded execution  provides strong security guarantees for legacy applications running on untrusted platforms~\cite{Baumann2014}. Prominent examples include Haven~\cite{Baumann2014}, SCONE~\cite{arnautov2016scone}, and Graphene-SGX~\cite{tsai2017graphene}.  Our work builds on the SCONE framework.
Intel SGX has become available in clouds~\cite{IBMCloudSGX, AzureSGX}, unleashing a plethora of services to be ported, including Web search~\cite{sgx-websearch},  actor framework~\cite{eactors}, storage~\cite{pesos, speicher}, leases~\cite{t-lease}, monitoring and profilers~\cite{tee-perf,teemon}, software update~\cite{TSR}, FaaS~\cite{clemmys}, networking~\cite{shieldbox, slick}, and data analytics systems~\cite{sgx-pyspark,opaque,Schuster2015}.



Recently, several secure machine learning systems have been proposed, which rely on Intel SGX to support secure machine learning~\cite{privado, slalom, chiron, ohrimenko}. 
Privado~\cite{privado} proposes a mechanism to obtain oblivious neural networks. Then, it executes the oblivious neural network inside SGX enclaves for secure inferencing.  Slalom~\cite{slalom} makes use of a combination of Intel SGX and untrusted GPUs to secure Deep Neural Networks (DNNs) computations. The idea of Slalom is that it splits the DNN computations into linear operations (e.g., matrix multiplications) on GPUs, whereas performing the non-linear operations  (eg. ReLUs operations) inside Intel SGX enclaves. This approach allows achieving much better performance since the intensive computation is performed with GPUs. Unfortunately, Slalom still has several limitations. First, as Privado, it focuses only on secure inferences. It refers to secure training computations as a research challenge. Second, it requires Tensorflow users to heavily modify or redevelop their existing code. Third, it does not support distributed settings, i.e., it does not support secure connections between SGX enclaves. Finally, Slalom is not production ready, in fact, it indicates that it can be used only for testing. 
Chiron~\cite{chiron} is the most relevant for \projectname, where they leveraged Intel SGX for privacy-preserving machine learning services. Unfortunately, Chiron is a single-threaded system within an enclave. In addition, Chiron requires adding an interpreter and model compiler into enclaves which introduce significant runtime overhead since the limited EPC size.  The work from Ohrimenko et al.~\cite{ohrimenko} also used Intel SGX to secure machine learning computations, however, it supports only a limited number of operators.  In contrast, we propose \projectname~ --- a practical distributed machine learning framework for securing both training and inference computations. 



\section{Conclusion} 
\label{sec:conclusion}
In this paper, we report on our experience with building and deploying \projectname,  a secure TensorFlow-based machine learning framework leveraging the hardware-assisted TEEs, specifically Intel SGX.   \projectname extends the  security properties of a secure stateless enclave in a single node to secure unmodified distributed stateful machine learning applications. Thereby, it provides a generic platform for end-to-end security for the input data, ML model, and application code. Moreover, it  supports both training and classification phases while providing all three important design properties for the secure machine learning workflow: {\em transparency},  {\em accuracy}, and {\em performance}. \projectname is a commercially available platform, and is currently being used in production by four major customers. While there are several open challenges and limitations of our system, our experience shows that \projectname strives for a promising approach: it incurs reasonable performance overheads, especially in the classification/inference process, while providing strong security properties against a powerful adversary.  Lastly, we also discussed several open challenges and on-going extensions to the system.

\myparagraph{Acknowledgements}
We thank our shepherd Professor Sara Bouchenak and the anonymous reviewers for their insightful comments and suggestions.
This work has received funding from the Cloud-KRITIS Project and the European Union's Horizon 2020 research and innovation programme under the LEGaTO Project (legato-project.eu), grant agreement No 780681.

\bibliographystyle{abbrv}
	\bibliography{main}

\end{document}